\documentstyle[prl,aps,preprint,psfig]{revtex}

\begin{document}
%@twocolumnfalse\endcsname
%%
%%
%\draft
%%
\title{Cosmological Constraint on the String Dilaton in 
Gauge-mediated Supersymmetry Breaking Theories}
\author{J. Hashiba}
\address{Department of Physics,  The University of
  Tokyo, Tokyo 113, Japan}
\author{M. Kawasaki}
\address{Institute for Cosmic Ray Research, The University of Tokyo,
  Tanashi 188, Japan}
\author{T. Yanagida}
\address{Department of Physics,  The University of
  Tokyo, Tokyo 113, Japan}
\date{\today}

\maketitle

\begin{abstract}
The dilaton field in string theories (if exists) is expected to have a mass 
of the order of the gravitino mass $m_{3/2}$ which is in a range of 
$10^{-2}$keV--1GeV in gauge-mediated supersymmetry breaking models.  If it 
is the case, the cosmic energy density of coherent dilaton oscillation 
easily exceeds the critical density of the present universe.  We show that 
even if this problem is solved by a late-time entropy production (thermal 
inflation) a stringent constraint on the energy density of the dilaton 
oscillation is derived from experimental upperbounds on the cosmic 
X($\gamma$)-ray backgrounds.  This excludes an interesting mass region, 
$500{\rm keV} \lesssim m_{3/2} \lesssim 1{\rm GeV}$, in gauge-mediated 
supersymmetry breaking models.

\end{abstract}


%]

The early universe has been used in the past couple of decades to study 
particle physics which often can not be tested directly by accelerator 
experiments.  If hypothetical particles predicted by particle physics exist 
in the universe, they may significantly affect the dynamical and thermal 
history of the universe.  In such a case, the present astrophysical 
observation gives stringent constraints on the abundances of the 
hypothetical particles and restricts the theory of particle physics.

The existence of a light dilaton field $\phi$ is one of generic
predictions in a large class of superstring theories~\cite{Green}. It
is expected to acquire its mass of the order of the gravitino mass
$m_{3/2}$ from some non-perturbative dynamics~\cite{Carlos}. In hidden
sector models for supersymmetry (SUSY) breaking the gravitino has a
mass at the electroweak scale. On the other hand, in gauge-mediated
SUSY breaking models~\cite{Dine,Dine2,Hotta,Poppitz,Murayama}, the
gravitino mass is predicted in a range of $10^{-2}$keV--1GeV. If the
dilaton has such a small mass $m_{\phi}\simeq m_{3/2}$, its lifetime
is much longer than the age of the present universe and the cosmic
energy density of its coherent oscillation easily exceeds the critical
density of the universe.

It has been pointed out in Ref.~\cite{Gauvea} that the above problem 
may be solved if a late-time thermal inflation~\cite{Lyth} takes 
place. Since the dilaton mass is very small $m_{\phi} \simeq 
10^{-2}$keV--1GeV, the thermal inflation seems only a mechanism to 
dilute substantially the cosmic energy density of the dilaton 
oscillation.\footnote{%%
If the dilaton has a mass of the order 10TeV as in a class of hidden 
sector models, we have no cosmological dilaton problem~\cite{Moroi}.} 

In this letter we show that a more stringent
constraint~\cite{Kawasaki} on the energy density of the dilaton
oscillation is derived from experimental upperbounds on the cosmic
X($\gamma$)-ray backgrounds.  We adopt the thermal
inflation  as a dilution mechanism of the cosmic dilaton
density.  The
obtained constraint\footnote{%%
A similar constraint has been derived without assuming a specific 
dilution mechanism~\cite{Kawasaki}. The previous constraint is 
weaker than the present one.}
excludes a region of $500{\rm keV}\lesssim m_{3/2}\lesssim 1{\rm GeV}$ 
which raises a new problem in a large class of gauge-mediated SUSY 
breaking models observed recently~\cite{Poppitz,Murayama}.

We briefly review the thermal inflation model proposed by Lyth and 
Stewart~\cite{Lyth}. The potential of the inflaton $\chi$ is given by
\begin{equation}
    V = V_{0} - m^{2}|\chi|^{2} + \frac{1}{M_{*}^{2n}}
    |\chi|^{2n+4}.
    \label{pot-inf}
\end{equation}
We suppose that the negative mass squared, $-m^{2}$, for $\chi$ is generated
by SUSY-breaking higher order corrections and assume it of the order
of the SUSY-breaking scale $m \simeq O(100)$GeV. The mass $M_{*}$ denotes
a cut-off scale of this effective theory.  We take $M_{*}$ as a free
parameter in
order to make a general analysis.\footnote{%%
The original thermal inflation model~\cite{Lyth} assumes $M_{*}$ to be
the gravitational scale $M_{G}=2.4\times 10^{18}$GeV.}
$V_{0}$ is set so that the cosmological constant vanishes. The 
vacuum expectation value of $\chi$ is given by
\begin{equation}
    \langle \chi \rangle \equiv M = 
    \left(\frac{1}{n+2}\right)^{\frac{1}{2(n+1)}}
    \left(mM_{*}^{n}\right)^{\frac{1}{n+1}},
    \label{chi-exp}
\end{equation}
and
\begin{equation}
    V_{0} = m^{2}M^{2} - \frac{M^{2n+4}}{M_{*}^{2n}}
    = \frac{n+1}{n+2}m^{2}M^{2}.
    \label{vacuum-erg}
\end{equation}
We use, hereafter, the vacuum-expectation value $M$ of $\chi$ instead 
of $M_{*}$ to characterize the potential of the inflaton.

The inflaton decay rate is important to estimate the reheating
temperature $T_{R}$. 
In the SUSY standard model we have only a possible renormalizable 
interaction of the inflaton $\chi$ with SUSY-standard model 
particles:
\begin{equation}
    W = \lambda \chi H \bar{H},
    \label{higgs-int}
\end{equation}
where $H$ and $\bar{H}$ are Higgs chiral supermultiplets. The coupling
constant $\lambda$ should be taken very small, $\lambda \lesssim
\mu_{H}/M$, so that the induced mass $\lambda\langle\chi\rangle =
\lambda M$ is at most the electroweak scale. Here, $\mu_{H}$ is the
SUSY-invariant mass for the Higgs multiplets. As pointed out in
Ref.~\cite{Lyth}, with this small coupling the decays of $\chi$ into a
pair of Higgs fields give rise to the reheating temperature high
enough ($T_{R} \gtrsim 10$MeV) to maintain the success of big bang
nucleosynthesis as far as $M \lesssim 10^{12}$GeV. To derive a
conservative constraint we assume the lowest value of the reheating
temperature $T_{R}=10$MeV in the present analysis, since lower $T_{R}$
yields weaker constraint as we will see later.

If the mass of $\chi$ is below the threshold of a pair production of 
Higgs fields, the above decay processes are, however, not energetically 
allowed.\footnote{%%
A possible decay into two pairs of bottom and antibottom quarks is
strongly suppressed by phase volume effects. We thank K.~Hikasa for
this point.}
Fortunately,  one-loop diagrams of the Higgs multiplets induce a 
coupling of $\chi$ to two photons as\footnote{%%
In the calculation we have neglected SUSY-breaking effects.}
\begin{equation}
    {\cal L}_{eff} = \frac{\alpha_{em}}{4\pi}
    \frac{\lambda}{\mu_{H}} \chi F_{\mu\nu}F^{\mu\nu}.
    \label{chi-gamma}
\end{equation}
We assume that the SUSY-invariant Higgs mass $\mu_{H}$ is dominated by 
the induced mass $\lambda M$, i.e. $\mu_{H} = \lambda M$. (We have a 
stronger constraint for $\lambda/\mu_{H} < 1/M$, otherwise.) The 
decay rate is given by
\begin{equation}
    \Gamma(\chi \rightarrow 2\gamma) = \frac{1}{4\pi}
    \left(\frac{\alpha_{em}}{4\pi}\right)^{2}
    \frac{m_{\chi}^3}{M^{2}},
    \label{gamma-decay}
\end{equation}
which leads to the reheating temperature
\begin{equation}
    T_{R} \simeq 0.25\frac{\alpha_{em}}{(4\pi)^{3/2}}
    \left(\frac{m_{\chi}}{M}\right)\sqrt{m_{\chi}M_{pl}}.
    \label{reheating-temp}
\end{equation}
Here, $M_{pl}$ is the Planck mass $M_{pl} =\sqrt{8\pi}M_{G}= 
1.2\times 10^{19}$GeV and $m_{\chi}$ the physical mass of inflaton $\chi$
around the vacuum $\langle\chi\rangle=M$. We find from 
Eqs.~(\ref{pot-inf}) and (\ref{chi-exp})
\begin{equation}
    m_{\chi}^{2} = 2(n+1)m^{2}.
    \label{flaton-mass}
\end{equation}

In summary, for the case of $m_{\chi} < 130$GeV\footnote{%%
The experimental lower bounds on the masses for Higgs bosons and Higgsinos 
are about 65GeV~\cite{LEP2}.}
we use the above reheating temperature (\ref{reheating-temp}) while
for the case of $m_{\chi}\ge 130$GeV we use the lowest possible value
for $T_{R}$, $T_{R} = 10$MeV, to derive a conservative constraint on
the cosmic
dilaton density.\footnote{%%
Since the potential Eq.(\ref{pot-inf}) possesses a global $U(1)$ 
symmetry, we have a massless Nambu-Goldstone (NG) boson (the angular 
part of $\chi$). The NG bosons produced by the $\chi$ decay are 
cosmologically dangerous~\cite{Lyth}. Thus, we assume that some 
explicit breaking of the $U(1)$ symmetry generates a mass of the NG 
boson large enough to suppress the $\chi$ decay into the NG
bosons. The detailed analysis with the explicit $U(1)$-breaking term
will be given in Ref.~\cite{Hashiba}.}

Let us now discuss dynamics of the thermal inflation and estimate the 
cosmic energy density of coherent dilaton oscillation in the 
present universe. 

When the inflaton $\chi$ couples to particles which are in 
thermal equilibrium,\footnote{%%
The particles coupled to $\chi$ have large masses of the order $M$ in
the true vacuum $\langle\chi\rangle = M$. But in the false vacuum
$\langle\chi\rangle \simeq 0$ they are light and could be in thermal bath.}
the effective potential of $\chi$ is written as 
\begin{equation}
        V_{eff} = V_{0} - m^{2}|\chi |^{2} + \frac{1}{M_{*}^{2n}}
        |\chi|^{2n+4}  +  cT^{2}|\chi|^2,
        \label{pot-inf-eff}
\end{equation}
where $T$ is the cosmic temperature and $c$ a constant of the order 1.
Then, at high temperature $T > T_{c} \simeq m$, the effective mass
squared of the inflaton $\chi$ is positive and $\chi$ sits near the
origin $\langle\chi\rangle \simeq 0$.  At this epoch the radiation
energy density is given by $\rho_{rad} = g_{*}\pi^{2}T^{4}/30$ where
$g_{*}$ is the effective number of degrees of freedom.  The important
fact is that $\rho_{rad}$ becomes less than the vacuum energy density
$V_0$ for $T < T_{*} \simeq V_{0}^{1/4}$.  Thus, at temperature
$T_{c} \lesssim T \lesssim T_{*}$ the vacuum energy $V_0$ dominates
the energy density of the universe and a mini-inflation (i.e. the
thermal inflation)
occurs~\cite{Lyth}.\footnote{%%
The energy density of the universe is most likely dominated by the
coherent dilaton oscillation just before the beginning of the thermal
inflation. In this case the thermal inflation starts at $T_{*} \simeq
(V_0^2/(m_{\phi}M_{G}))^{1/6}$, which does not, however, affect the
present analysis as long as $T_{*} > T_{c}$.}

After $T < T_{c}$ the effective mass squared of the inflaton $\chi$
becomes negative and $\chi$ rolls down toward the true minimum
(\ref{chi-exp}) of the potential (\ref{pot-inf}) and the thermal
inflation ends. Then, the inflaton $\chi$ oscillates around it.  The
energy density of the oscillating field is finally transferred to the
radiation with temperature $T_{R}$ through the inflaton decay.
Therefore, the thermal inflation increases the entropy of the universe
by a factor
\begin{equation}
        \Delta \simeq \frac{4V_{0}/3T_{R}}
        {(2\pi^2/45)g_{*}T_{c}^{3}}
        \simeq \frac{V_{0}}{70T_{R}T_{c}^{3}}.
        \label{dilution}
\end{equation}
Here we have used the fact that the energy density of the oscillating
field and the entropy density decrease as $a^{-3}$ where $a$ is the
scale factor of the universe.

The dilaton $\phi$ starts to oscillate with the initial
amplitude $\phi_{0}$ when its mass $m_{\phi}$ becomes larger than the
Hubble parameter $H$.  Hereafter, we call this dilaton as `big bang
dilaton'.  In order for the thermal inflation to dilute this dilaton
energy density the dilaton $\phi$ should start to oscillate before the
thermal inflation. Thus we assume that the Hubble parameter $H$ during
the thermal inflation is less than $m_{\phi}$.  Then, the
abundance of the `big bang dilaton' before the thermal inflation is
given by~\cite{Lyth}\footnote{%%
We have assumed that full reheating after ordinary inflation occurs
before the thermal inflation. A more detailed analysis without this
assumption will be given by in Ref.~\cite{Hashiba}}
\begin{equation}
    \left(\frac{n_{\phi}}{s}\right)_{BB} \simeq
    \frac{m_{\phi}\phi_{0}^{2}/2}{8.6m_{\phi}^{3/2}M_{G}^{3/2}}
    \simeq \frac{\phi_{0}^{2}}{17m_{\phi}^{1/2}M_{G}^{3/2}}.
    \label{bb-moduli-initial}
\end{equation}
The abundance of `big bang dilaton' after the thermal inflation is given by 
\begin{equation}
    \left(\frac{n_{\phi}}{s}\right)_{BB} \simeq 4
    \left(\frac{T_{c}}{m_{\chi}}\right)^{3}
    \left(\frac{\phi_{0}}{M_{G}}\right)^{2}
    \left(\frac{M_{G}}{m_{\phi}}\right)^{1/2}
    \frac{m_{\chi}^{3}T_{R}}{V_{0}}.
    \label{bb-moduli}
\end{equation}

In addition to the `big bang dilaton', the dilaton oscillation is
reproduced by the thermal inflation since the minimum of the potential
is shifted from its true vacuum by an amount of $\delta \phi \sim
(V_{0}/m_{\phi}^2M_{G}^2)\phi_{0}$~\cite{Lyth}, which results in the
dilaton density $\sim m_{\phi}^{2}\delta\phi^{2}/2$.  The abundance of
this `thermal inflation dilaton' is estimated as
\begin{equation}
    \left(\frac{n_{\phi}}{s}\right)_{TI} \simeq 
    \frac{3}{8} \left(\frac{\phi_{0}}{M_{G}}\right)^{2}
    \frac{V_{0}T_{R}}{m_{\phi}^{3}M_{G}^{2}}.
    \label{ti-moduli}
\end{equation}

Let us consider first the case of $m_{\chi}\ge 130$GeV. 
The lower bound of total energy density of the dilaton $\phi$ is given by 
\begin{eqnarray}
    \frac{\rho_{\phi}}{s} & \simeq & m_{\phi}
    \max \left[\left(\frac{n_{\phi}}{s}\right)_{BB},
      \left(\frac{n_{\phi}}{s}\right)_{TI}\right]
    \gtrsim m_{\phi}\sqrt{\left(\frac{n_{\phi}}{s}\right)_{BB}
    \left(\frac{n_{\phi}}{s}\right)_{TI}}
    \nonumber  \\
    & = & \sqrt{\frac{3}{2}}
    \left(\frac{\phi_{0}}{M_{G}}\right)^{2}
    \left(\frac{T_{c}}{m_{\chi}}\right)^{3/2}
    \left(\frac{m_{\chi}^{2}}{m_{\phi}M_{G}}\right)^{3/4}T_{R}.
    \label{min-density}
\end{eqnarray}
We see that the minimum abundance is given for the lowest possible
reheating temperature $T_{R} = 10$MeV. For $\phi_{0}\simeq M_{G}$,
$T_{c}\simeq m_{\chi}$, $T_{R}\simeq 10$MeV and $m_{\chi} \ge 130$GeV
we obtain the lower abound of $\rho_{\phi}/s$ as
\begin{equation}
    \frac{\rho_{\phi}}{s} \gtrsim 5.3\times 10^{-11}{\rm GeV}
    \left(\frac{m_{\phi}}{\rm MeV}\right)^{-3/4}.
\end{equation}
Comparing with the critical density of the present universe,
$\rho_{c}/s = 3.6\times 10^{-9}h^{2}{\rm GeV}$, we find
\begin{equation}
    \Omega_{\phi}h^{2}\equiv \frac{\rho_{\phi}h^{2}}{\rho_{c}}
    \gtrsim 1.5\times 10^{-2}
    \left(\frac{m_{\phi}}{\rm MeV}\right)^{-3/4},
    \label{omega-dilaton}
\end{equation}
where $h$ is the present Hubble parameter in units of 100km/sec/Mpc. 
We show in Fig.~1 this lower bound. We see that the predicted lower 
bound of the dilaton density may be taken below the critical 
density for $m_{\phi} \gtrsim 20$keV. 

Now we consider the case of $m_{\phi} < 130$GeV. In this case we use
the reheating temperature $T_{R}$ given in Eq.~(\ref{reheating-temp})
to write $M$ in terms of $T_{R}$ and $m_{\chi}$ as $M \simeq
10^{-4}m_{\chi}^{3/2}M_{G}^{1/2}T_{R}^{-1}$.  
Then $V_{0}$ is written from Eqs.~(\ref{vacuum-erg}) 
and (\ref{flaton-mass}) as $V_{0}\simeq
10^{-8}(2n+4)^{-1} m_{\chi}^{5}M_{G}T_{R}^{-2}$. The abundance of the
dilaton are given by
\begin{equation}
    \left(\frac{n_{\phi}}{s}\right)_{BB} \simeq 4
    \left(\frac{T_{c}}{m_{\chi}}\right)^{3}
    \left(\frac{\phi_{0}}{M_{G}}\right)^{2}
    \left(\frac{M_{G}}{m_{\phi}}\right)^{1/2}
    \frac{(2n+4)10^{8}T_{R}^{3}}{m_{\chi}^{2}M_{G}}.
    \label{bb-moduli2}
\end{equation}
and 
\begin{equation}
    \left(\frac{n_{\phi}}{s}\right)_{TI} \simeq 
    \frac{3}{8} \left(\frac{\phi_{0}}{M_{G}}\right)^{2}
    \frac{10^{-8}m_{\chi}^{5}}{(2n+4)m_{\phi}^{3}M_{G}T_{R}}.
    \label{ti-moduli2}
\end{equation}
The lower bound of the total abundance is achieved when
$(n_{\phi}/s)_{BB} = (n_{\phi}/s)_{TI}$, namely
\begin{equation}
    m_{\chi, min}\simeq 190 
    (2n+4)^{2/7}M_{G}^{1/14}T_{R}^{4/7}m_{\phi}^{5/14}.
    \label{chi-min}
\end{equation}
This yields (from Eq.(\ref{min-density}))
\begin{equation}
    \Omega_{\phi}h^{2} \gtrsim 1.5 \times 10^{-2}
    \left(\frac{m_{\chi,min}}{130{\rm GeV}}\right)^{3/2}
    \left(\frac{m_{\phi}}{\rm MeV}\right)^{-3/4} 
    \simeq 2.3\times 10^{-3}
    \left(\frac{m_{\phi}}{\rm MeV}\right)^{-3/14}.
\end{equation}
Here we have taken $(2n+4)^{3/7}\simeq 2$, $\phi_{0}\simeq
M_{G}$, $T_{c}\simeq m_{\chi}$ and $T_R \simeq 10$MeV. We also show
this lower bound in Fig.~1. Notice that a kink appears at $m_{\phi}
\simeq 20$MeV. This is because $m_{\chi, min}$ in Eq.~(\ref{chi-min})
exceeds 130GeV for $m_{\phi} \simeq 20$MeV and the dilaton abundance
takes its minimum at $m_{\chi}= 130$GeV. We see from Fig.~1 that for
all region of $m_{\phi}\simeq 10^{-2}{\rm keV} - 1{\rm GeV}$ relevant
to gauge-mediated SUSY breaking models the lower bound of
$\Omega_{\phi}h^{2}$ is below the critical density $\Omega h^{2}
\simeq 0.25$ in the present universe.

We are now at the point to derive a constraint from the observed 
X($\gamma$)-ray backgrounds.

First, we should estimate the lifetime of the dilaton. The main decay
mode of dilaton is a two-photon process, $\phi \rightarrow 2\gamma$,
since the decay mode to two neutrinos has a chirality suppression and
vanishes for massless neutrinos~\cite{Kawasaki}. The dilaton $\phi$
has a couplings to two photons as
\begin{equation}
    {\cal L}_{\phi} = \frac{b}{16\pi\alpha_{em}}
    \frac{\phi}{M_{G}}F_{\mu\nu}F^{\mu\nu}.
    \label{dilaton-photon}
\end{equation}
Here, $b$ is a parameter of the order 1 which depends on the type of 
superstring theories and compactifications.\footnote{%%
For example we get $b=\sqrt{2}$~\cite{Dudas} for a compactification of
the M-theory~\cite{Horava}. }
We simply take $b=1$ in the present analysis.\footnote{%%
We note that our conclusion does not depend heavily on the parameter
$b$. If one takes $b=0.3$, for instance, the excluded region ($500{\rm
keV} \lesssim m_{\phi} \lesssim 1{\rm GeV}$) derived in the present
analysis moves slightly to $1{\rm MeV} \lesssim m_{\phi} \lesssim
1{\rm GeV}$.}
Then, the lifetime 
$\tau_{\phi}$ of the dilaton is estimated as 
\begin{equation}
    \tau_{\phi} \simeq 6\frac{M_{pl}^{2}}{m_{\phi}^{3}}
    \simeq 7\times 10^{23}\sec
    \left(\frac{m_{\phi}}{\rm MeV}\right)^{-3}.
    \label{lifetime}
\end{equation}
The X($\gamma$)-ray flux from the dilaton decay is given 
by~\cite{Kawasaki} 
\begin{equation}
    F_{\gamma}(E) = 
    \frac{n_{\phi,0}}{2\pi\tau_{\phi}H_{0}\Omega_{0}}
    \left(\frac{E}{m_{\phi}}\right)^{3/2}
    \exp\left[-\frac{2}{3\tau_{\phi}H_{0}\Omega_{0}^{1/2}}
      \left(\frac{E}{m_{\phi}}\right)^{3/2}\right],
    \label{flux}
\end{equation}
where $n_{\phi,0}$ is the present dilaton number density, $H_{0}$ the
present Hubble parameter, $\Omega_{0}$ the total density of the
present universe in units of the critical density, and $E$ the energy
of X($\gamma$)-ray. The flux $F_{\gamma}$ takes the  maximum value
$F_{\gamma,max}$ at
\begin{eqnarray}
    E_{max}  & = &
          \frac{m_{\phi}}{2}   
          ~~~~~~~~~~~~~~~~~~~~~~~~~~~~~~~{\rm for}~~\tau_{\phi} > 
          \frac{2}{3}H_{0}^{-1}\Omega_{0}^{-1/2} \nonumber \\
          & = & \frac{m_{\phi}}{2}
          \left(\frac{3\tau_{\phi}H_{0}\Omega_{0}^{1/2}}{2}\right)^{2/3} 
          ~~~~~~~~{\rm for}~~\tau_{\phi} <
          \frac{2}{3}H_{0}^{-1}\Omega_{0}^{-1/2} .
\end{eqnarray}
By requiring that $F_{\gamma,max}$ should be less than the observed
X($\gamma$)-ray backgrounds~\cite{Gendreau,HEAO,Kappadath}, we obtain
a constraint on $\Omega_{\phi}h^{2}$ which is also shown in Fig.~1. We see
that the mass region $500{\rm keV} \lesssim m_{\phi} \lesssim
1{\rm GeV}$ is excluded by the observed X($\gamma$)-ray backgrounds.

In summary, we have shown that generic models for the thermal inflation
are successful to dilute the energy density of
the coherent dilaton oscillation below the critical density of the
present universe. However, we have found that the constraint from the
experimental upperbounds on the cosmic X($\gamma$)-ray backgrounds is
much more stringent and it excludes the dilaton mass region, $500{\rm keV}
\lesssim m_{\phi} \lesssim 1{\rm GeV}$.\footnote{%%
If one assumes  $\phi_0 \simeq 0.01M_G$, a small window around
$m_{\phi}\simeq 1$GeV appears.}
This raises a new problem in recently observed interesting
models~\cite{Hotta,Poppitz} for gauge-mediated SUSY breaking as long
as $m_{\phi} \simeq m_{3/2}$ as expected in a large class of
superstring theories. On the other hand, the region, $10^{-2}{\rm keV}
\lesssim m_{\phi} \lesssim 500{\rm keV}$, survives the constraint. In
this region the lower bounds of the dilaton density $\Omega_{\phi}h^2$
are achieved when $m_{\chi}\simeq 10$GeV and $M\simeq 10^9$GeV which
implies the cut-off scale $M_{*}\simeq 10^{17}$GeV for $n=1$. It may be
encouraging that the cut-off scale $M_{*}$ is not far below the
gravitational scale $M_{G}\simeq 2.4\times 10^{18}$GeV.

Two of the authors (J.H. and T.Y.) thank K. Hikasa, K. Hosomichi,
Y. Nomura and K. Tobe for useful comments.

\begin{figure}
    \centerline{\psfig{figure=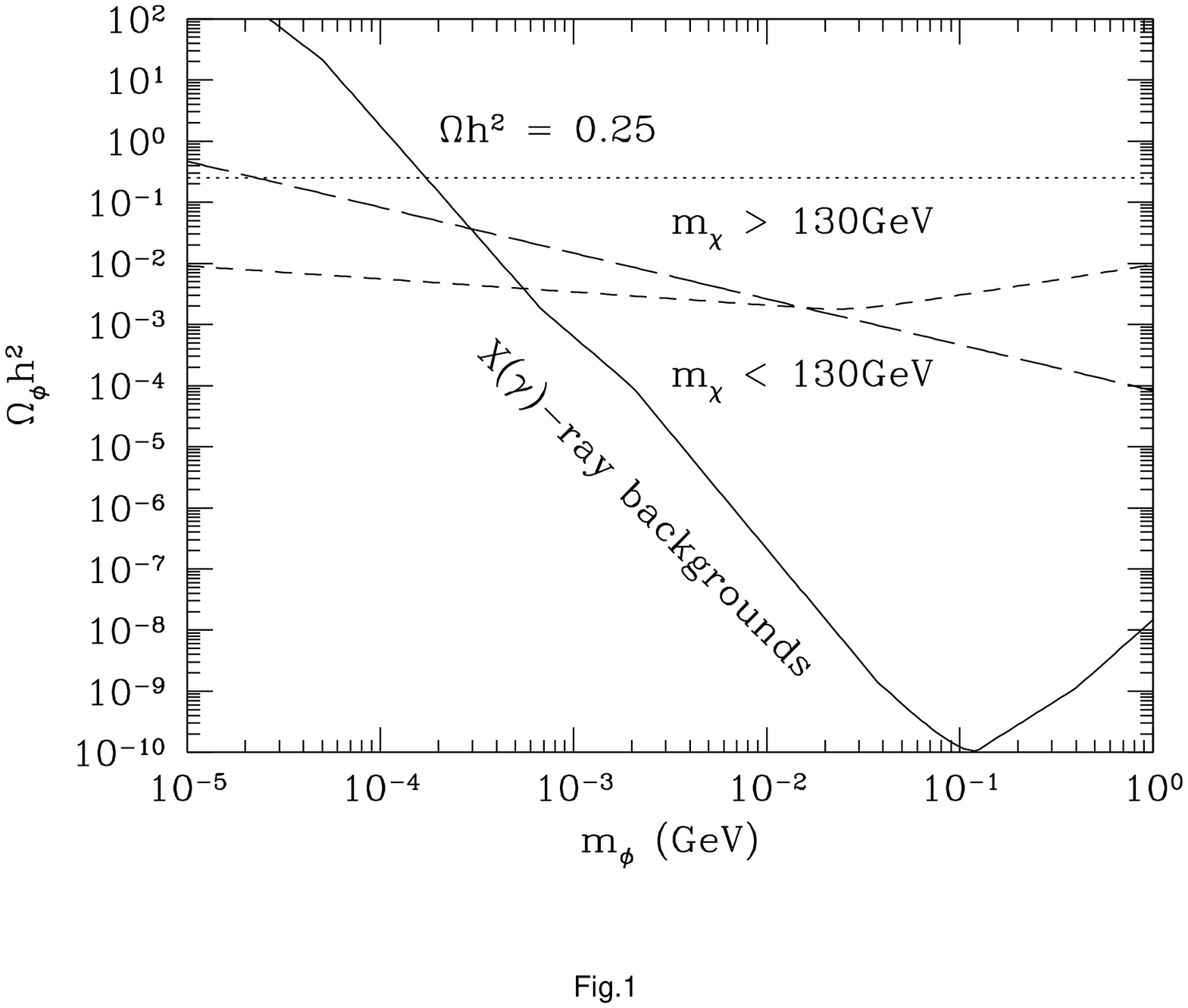,width=14cm}}
    \vspace{0.5cm}
    \caption{The lower bounds of the dilaton density
    $\Omega_{\phi}h^2$ in the presence of the thermal inflation for
    $m_{\phi} > 130$GeV (long-dashed line) and $m_{\phi} < 130$GeV
    (short-dashed line).  The dotted line represents $\Omega h^2 =
    0.25$. The experimental upperbound on $\Omega_{\phi}h^2$ from the
    X-ray backgrounds is shown by solid curve.}
\end{figure}

\end{document}